\documentclass[12pt]{iopart}
\usepackage{graphics}
\usepackage{epsfig}
\usepackage{iopams}
\usepackage{setstack}
\newcommand{\fr}{{^F\hspace{-.02in}R}}

\def\eqref#1{(\ref{#1})}
\def\diag{\mathop{\rm diag}}
\begin{document}

\title[]{Ultrarelativistic Motion: Inertial and Tidal Effects in
Fermi Coordinates}

\author{C. Chicone\dag, B. Mashhoon\ddag}

\address{\dag\ Department of Mathematics, University of
Missouri-Columbia, Columbia, Missouri 65211, USA}
\address{\ddag\ Department of Physics and
Astronomy, University of Missouri-Columbia, Columbia, Missouri 65211, USA }

\ead{MashhoonB@missouri.edu}

\begin{abstract}
Fermi coordinates are the natural generalization of inertial 
Cartesian coordinates to accelerated systems and gravitational fields. 
We study the motion of ultrarelativistic particles and light rays 
in Fermi coordinates and investigate inertial and tidal effects 
beyond the critical speed $c/\sqrt{2}$. In particular, we discuss the 
black-hole tidal acceleration mechanism for ultrarelativistic particles 
in connection with a possible origin for high-energy cosmic rays.
\end{abstract}

%\submitto{\CQG}

\maketitle

\section{Introduction}\label{s1}
Physically meaningful interpretation of the measurement of relative motion 
from the point of view of an accelerated observer in a gravitational field
requires the introduction
of a special coordinate system (i.e.\ Fermi coordinates) along the 
worldline of the observer. In this coordinate system,  the equations
of relative motion reveal tidal and inertial effects for ultrarelativistic
motion (with speed exceeding the critical value $c/\sqrt{2}$) that are
contrary to Newtonian expectations. 
While there are general treatments of inertial and tidal effects
in Fermi coordinates~\cite{1,2,3,4} 
and the special case of ultrarelativistic motion of particles
has been discussed in our recent papers~\cite{5,new6,6},  the purpose of
this work is to present a more systematic and complete description
of motion beyond the critical speed $c/\sqrt{2}$. To this end,
the equations of motion in Fermi coordinates are discussed in section~\ref{s0}
and inertial effects are considered in sections~\ref{s2}, \ref{s3} 
and appendix A. Section~\ref{s4} is devoted to tidal effects. A summary and
brief discussion of our results is contained in section~\ref{s5}.

\section{Equations of motion in Fermi coordinates}\label{s0}  

Imagine an accelerated observer in a general relativistic spacetime
following a worldline $\bar{x}^\mu (\tau )$, where $\tau$ is the
proper time along its trajectory. 
The local axes of the observer are
given by an orthonormal tetrad frame $\lambda^\mu _{\;\; (\alpha )}$
that is carried along its path according to
\begin{equation}\label{eq1}
\frac{D\lambda^\mu_{\;\;(\alpha )}}{{\rm d}\tau }=\Phi _{(\alpha
)}^{\;\;\;\; (\beta )}\lambda^\mu_{\;\;(\beta)},
\end{equation}
where $\Phi_{(\alpha )(\beta)}$ is the antisymmetric acceleration
tensor, $\lambda^\mu_{\;\; (0)}={\rm d}\bar{x}^\mu /{\rm d}\tau$ 
is the local
temporal direction and $\lambda^\mu_{\;\;(i)}$, $i=1,2,3$, form the
local spatial triad, and where (here and throughout this paper) 
Greek indices run from 0
to 3, Latin indices run from 1 to 3, the signature of the metric is
$+2$ and $c=1$, unless specified otherwise. 

In analogy with the
electromagnetic Faraday tensor, we decompose the acceleration tensor
into its ``electric'' and ``magnetic'' components $\phi_{(\alpha
)(\beta)}\to (-{\bf a},\boldsymbol{\omega})$; these are given by the
translational acceleration ${\bf a}(\tau )$ and the rotational
frequency $\boldsymbol{\omega}(\tau )$, respectively. That is, the
acceleration of the reference trajectory is given by
$a^i\lambda^\mu_{\;\;(i)}$ and $\boldsymbol{\omega}$ is the frequency
of rotation of the spatial triad with respect to a local nonrotating
(i.e.\ Fermi-Walker transported) triad.

Let us next establish a Fermi coordinate system $(T,{\bf X})$ in a
neighborhood of the reference worldline~\cite{1}.
It turns out that Fermi coordinates can be assigned uniquely 
only to spacetime events that are within a cylindrical region of 
finite radius along the observer's worldline. At each event
$\bar{x}^\mu (\tau )$ on the observer's worldline, consider all
spacelike geodesic curves that are normal to this 
reference worldline at this event. Each event $x^\mu$  
on the resulting 
hypersurface and within the 
cylindrical region under consideration is connected to
$\bar{x}^\mu (\tau )$ by a unique spacelike geodesic curve that starts
at $\bar{x}^\mu (\tau )$ and whose tangent
vector $\xi ^\mu$  at this event is normal to 
the reference worldline (that is, $\xi_\mu \lambda^\mu_{\;\;(0)} =0$).
The event $x^\mu$ is assigned the 
Fermi coordinates $X^\mu =(T,{\bf X})$, where
$T=\tau $ and 
\begin{equation}\label{eq2}
X^i=\sigma \xi ^\mu \lambda_\mu ^{\;\;(i)}.
\end{equation}
Here $\sigma$ is the proper length of the normal spacelike geodesic segment
connecting  $\bar{x}^\mu (\tau )$ to $x^\mu$. The Fermi coordinate
system is admissible in a cylindrical spacetime region around
$\bar{x}^\mu (\tau )$ with radius $|{\bf X}|\sim \mathcal{R}$, where
$\mathcal{R}$ is the minimum radius of curvature of spacetime along
the reference worldline~\cite{2}.

The observer, at each instant $\tau =T$ of its proper time ``sees'' a
locally Euclidean three-dimensional space and ``instantaneously''
determines distances within it using the (spacelike) geodesic lengths
beginning from its position at the spatial origin of the new Fermi
coordinates. A nearby particle worldline that punctures the sequence
of three-dimensional spaces at ${\bf X}=(X(T),Y(T),Z(T))$ is thus a
graph over the reference worldline and each point on the graph 
is connected to the observer
by a ``locally straight''  line (i.e.\ a geodesic) that is 
normal to the observer's worldline. From the viewpoint of the observer,
the particle has  
relative coordinate
velocity  ${\bf V}={\rm d}{\bf X}/{\rm d}T$ and  relative
coordinate acceleration  ${\rm d}{\bf V}/{\rm d}T$.

The general equation of motion for a test particle of mass $m$ in the
Fermi frame is 
\begin{equation}\label{eq3}
\frac{{\rm d}^2X^\mu}{{\rm d}s^2}+\Gamma^\mu_{\alpha \beta}\frac{{\rm d}X^\alpha}{{\rm d}s}
\frac{{\rm d}X^\beta}{{\rm d}s}=A^\mu ,
\end{equation}
where $s$ is the proper time along its worldline, 
$-{\rm d}s^2=g_{\mu\nu}(T,{\bf X}){\rm d}X^\mu {\rm d}X^\nu$, and $mA^\mu =F^\mu$ is the external
force acting on the particle. 
 
Equation~\eqref{eq3} can be written as the system
\begin{eqnarray}
\label{eq4} \frac{{\rm d}^2T}{{\rm d}s^2}+\Gamma^0_{\alpha
\beta}\frac{{\rm d}X^\alpha}{{\rm d}s} \frac{{\rm d}X^\beta}{{\rm d}s}=A^0,\\
\label{eq5} \frac{{\rm d}^2X^i}{{\rm d}s^2}+\Gamma^i_{\alpha \beta}
\frac{{\rm d}X^\alpha}{{\rm d}s} \frac{{\rm d}X^\beta}{{\rm d}s}=A^i.
\end{eqnarray}
Using the identity
\begin{equation}\label{eq6}
\frac{{\rm d}^2X^i}{ds^2}
=\frac{{\rm d}^2T}{{\rm d}s^2}V^i+\Gamma ^2\frac{{\rm d}^2X^i}{{\rm d}T^2},
\end{equation}
and the Lorentz factor of the particle
\begin{equation}\label{eq8}
\Gamma (T,{\bf X}):=\frac{{\rm d}T}{ds} ,
\end{equation}
equation~\eqref{eq5} can be expressed as
\begin{equation}\label{eq7}
\frac{{\rm d}^2X^i}{{\rm d}T^2} +(\Gamma^i_{\alpha \beta}-\Gamma^0_{\alpha
\beta}V^i)\frac{{\rm d}X^\alpha}{{\rm d}T} \frac{{\rm d}X^\beta}{{\rm d}T}=\frac{1}{\Gamma^2}
(A^i-A^0V^i).
\end{equation}

Let $U^\mu :={\rm d}X^\mu /{\rm d}s$. We note that 
$U^\mu =\Gamma (1,{\bf V})$ and 
the physical content of the equation of motion~\eqref{eq3} is
contained in equation~\eqref{eq7} together with $U_\mu U^\mu =-1$ and
$U_\mu A^\mu =0$. These can be written respectively as
\begin{equation}\label{eq9}
\Gamma = \frac{1}{\sqrt{-g_{00}-2g_{0i}V^i-g_{ij}V^iV^j}}
\end{equation}
and
\begin{equation}\label{eq10}
A^0=-\frac{g_{0j}+g_{ij}V^i}{g_{00}+g_{0i}V^i} A^j.
\end{equation}
It is crucial to recognize that in Fermi coordinates 
the velocity ${\bf V}$ satisfies the condition
$|{\bf V}| \le 1$ only at ${\bf X} = 0$; 
indeed, away from the reference trajectory $|{\bf V}|$
could in principle exceed unity in accordance with the equation of 
motion~\eqref{eq7}.

\section{Inertial effects}\label{s2}
It is natural to begin our discussion with an accelerated observer in
Minkowski spacetime. In this case, spacelike geodesic segments are 
straight lines; therefore, the construction of Fermi coordinates is simple.
In fact, equation~\eqref{eq2} reduces to $x^\mu -\bar{x}^\mu(\tau
)=X^i\lambda^\mu_{\;\; (i)}(\tau )$. Differentiating this relation we
find that
\begin{equation}\label{eq11}
{\rm d}x^\mu = [P\lambda^\mu_{\;\; (0)}+Q^j\lambda^\mu
_{\;\;(j)}]{\rm d}X^0+\lambda^\mu _{\;\;(i)} {\rm d}X^i,
\end{equation}
where $P$ and $Q$ are given by
\begin{equation}\label{eq12}
P(T,{\bf X})=1+{\bf a}(T)\cdot {\bf X}, \quad {\bf Q}(T,{\bf
X})=\boldsymbol{\omega} (T)\times {\bf X}.
\end{equation}
The components $g_{\mu\nu}$ of the Minkowski metric tensor 
% $-{\rm d}s^2=\eta_{\alpha \beta}{\rm d}x^\alpha {\rm d}x^\beta$ 
% when transformed to 
in Fermi coordinates are given by
\begin{equation}\label{eq13}
g_{00} =-P^2+Q^2,\quad g_{0i}=Q_i,\quad g_{ij}=\delta_{ij}.
\end{equation}
Moreover,  $\det (g_{\mu\nu})=-P^2$ and the inverse
metric is given by
\begin{equation}\label{eq14}
g^{00}=-\frac{1}{P^2},\quad g^{0i}=\frac{Q_i}{P^2},\quad
g^{ij}=\delta_{ij}-\frac{Q_iQ_j}{P^2}.
\end{equation}
The Christoffel symbols can be evaluated using \eqref{eq13} and
\eqref{eq14}; the nonzero components are
\begin{eqnarray}
\label{eq15} \Gamma^0_{00}=\frac{{\bf S}\cdot {\bf X}}{P},\quad
\Gamma^0_{0i}=\frac{a_i}{P},
\quad \Gamma^i_{0j}=-\big( \epsilon_{ijk}\omega^k+\frac{Q_ia_j}{P}\big),\\
\label{eq16} \Gamma^i_{00}=Pa_i -\frac{{\bf S}\cdot {\bf
X}}{P}Q_i+[\boldsymbol{\omega}\times (\boldsymbol{\omega}\times {\bf
X})+\dot{\boldsymbol{\omega}}\times {\bf X}]_i,
\end{eqnarray}
where an overdot denotes differentiation with respect to $T$ and ${\bf
S}$ is defined by
\begin{equation}\label{eq17}
{\bf S}(T):=\dot{{\bf a}}+{\bf a}\times \boldsymbol{\omega}.
\end{equation}

In Minkowski spacetime, the Fermi coordinates are admissible for $P^2>Q^2$. 
We note that the structure of the boundary region $P^2=Q^2$ has been discussed in
detail~\cite{7}. In general, the boundary of the admissible region
at a given time $T$ is a real quadric cone. If $\boldsymbol{\omega} =0$, this
surface degenerates into coincident planes. Moreover, 
if $\boldsymbol{\omega} \neq 0$, but 
${\bf a}\cdot \boldsymbol{\boldsymbol{\omega}}=0$, the boundary surface is
a hyperbolic cylinder for $|\boldsymbol{\omega}|^2<|{\bf a}|^2$, a parabolic cylinder for
$|\boldsymbol{\omega}| ^2=|{\bf a}|^2$, and an elliptic cylinder for 
$|\boldsymbol{\omega}|^2>|{\bf a}|^2$. As is
well known, for ${\bf a}=0$, the boundary surface is a circular cylinder of
radius $|\boldsymbol{\omega}|^{-1}$.

We now consider the equations of motion in Fermi
coordinates. Equations~\eqref{eq9} and~\eqref{eq10} are given by
\begin{eqnarray}
\label{eq18} \frac{1}{\Gamma^2} = P^2-({\bf Q}+{\bf V})^2\geq 0,\\
\label{eq19} A^0=\frac{({\bf Q}+{\bf V})\cdot {\bf A}}{P^2 -Q^2-{\bf
Q}\cdot {\bf V}}\;.
\end{eqnarray}
Hence, equation~\eqref{eq7} has the form
\begin{eqnarray}\nonumber
\fl \frac{{\rm d}^2 {\bf X}}{{\rm d}T^2}+ 2\boldsymbol{\omega} \times {\bf
V}+\boldsymbol{\omega}\times (\boldsymbol{\omega}\times {\bf
X})+\dot{\boldsymbol{\omega}}\times {\bf X}\\
+ P{\bf
a}-\frac{1}{c^2P}({\bf Q}+{\bf V})({\bf S}\cdot {\bf X}+2{\bf a}\cdot
{\bf V})=\boldsymbol{\mathcal{F}},
\label{eq20}\end{eqnarray}
where $\boldsymbol{\mathcal{F}}$ represents the external force per
unit mass
\begin{equation}\label{eq21}
\boldsymbol{\mathcal{F}}=\frac{1}{\Gamma^2}\big [\, {\bf A}
 -\frac{({\bf Q}+{\bf V})\cdot {\bf A}}
       {{c^2(P^2-Q^2-{\bf Q}\cdot {\bf V})}} 
{\bf V}\,\big] .
\end{equation}
It is interesting to note that in equation~\eqref{eq20} the
purely rotational inertial accelerations are essentially the same
as in the nonrelativistic theory~\cite{new,6}.

Inertial accelerations have been discussed by a number of 
authors~\cite{3,4,6,8,9,10,11,12,13}. 
In particular, it has been shown~\cite{6} that
for $\boldsymbol{\omega}=0$ and $\dot{{\bf a}}=0$, the inertial
acceleration experienced by the particle parallel to its motion is
given to lowest order in ${\bf a}$ by $-{\bf a}\cdot \hat{{\bf V}}
(1-2V^2/c^2)$, where ${\hat {\bf V}} = {\bf V}/ V$ 
is the unit vector tangent to the spatial path of the particle; 
therefore, there is a sign reversal for
$V>V_c=c/\sqrt{2}$ with consequences that are contrary to Newtonian
expectations.

Note that the acceleration ${\bf A}$ and the
inertial acceleration ${\bf a}$ occur in rather different ways in
equation~\eqref{eq20}. This difference is a consequence of the absolute
character of acceleration in the theory of relativity; that is, the fact
that an observer is accelerated is independent of the choice of coordinates. 
In particular, while the critical speed associated with ${\bf A}$ is $c$ 
(on the basis of equation~\eqref{eq21}), the  critical speed associated
with ${\bf a}$ in equation~\eqref{eq20} is
$c/\sqrt{2}$.

To obtain the limiting case of lightlike motion, we let ${\rm d}s=m\,{\rm d}\lambda$,
where $\lambda$ is an affine parameter. In the limit as $m\to 0$,  we have
that $P^2=({\bf Q}+{\bf V})^2$; hence, equation~\eqref{eq20}
is valid with $\boldsymbol{\mathcal{F}}=0$ (see~appendix A).

\section{Rindler observer}\label{s3}
For a Rindler observer~\cite{14} (that is, an observer in
hyperbolic motion in Minkowski spacetime), 
consider an inertial frame in which the Rindler observer is at rest at
$(0,0,0,z_0)$ and has uniform translational acceleration $g$ along the
$z$-axis. The observer's worldline is given by
\begin{equation}\label{eq22}
\bar{t}=\frac{1}{g}\sinh g\tau,\quad \bar{x}=\bar{y}=0,\quad
\bar{z}=z_0+\frac{1}{g}(-1+\cosh g\tau ).
\end{equation}
The nonrotating orthonormal tetrad frame along 
this worldline has the form
\begin{eqnarray}
\label{eq23} \lambda^\mu_{\;\;(0)} =\gamma (1,0,0,\beta ),\qquad
& \lambda ^\mu_{\;\;(1)}=(0,1,0,0),\\
\label{eq24} \lambda^\mu_{\;\;(2)} =(0,0,1,0), &
\lambda^\mu_{\;\;(3)}=\gamma (\beta ,0,0,1),
\end{eqnarray}
where $\beta =\tanh g\tau$ and $\gamma =\cosh g\tau$, and the
transformation from Fermi to inertial coordinates is
\begin{eqnarray}
\label{eq25} t=\Big ( Z+\frac{1}{g}\Big ) \sinh gT, \qquad x=X,\quad
y=Y,\\
\label{eq26}z=z_0-\frac{1}{g}+\Big(Z+\frac{1}{g}\Big)\cosh gT.
\end{eqnarray}
The Rindler coordinates are admissible for $T,X,Y\in (-\infty,\infty)$ 
and $Z\in (-1/g,\infty )$. We note that the manifold where $Z=-1/g$ is the 
Rindler horizon. It is related to the null cone in the inertial frame, since
equations~\eqref{eq25} and~\eqref{eq26} imply that
$(Z+1/g)^2=(z-z_0+1/g)^2-t^2$.

Imagine a free particle moving along the $z$-axis with speed $v_0$,
$-1<v_0<1$, according to
\begin{equation}\label{eq27}
z=v_0t+z_0.
\end{equation}
{}From the viewpoint of the Rindler observer, this motion is given by
\begin{equation}\label{eq28}
Z=\frac{1}{g}\Big(-1+\frac{1}{\cosh gT-v_0\sinh gT}\Big).
\end{equation}
For $v_0<0$, the particle descends monotonically toward the
horizon. For $v_0>0$, its path crosses the path of the Rindler
observer $(Z=0)$ at two events corresponding to $T=t=0$ and $T=T_1$, where
\begin{equation}\label{eq29}
T_1=\frac{1}{g}\ln \Big( \frac{1+v_0}{1-v_0}\Big).
\end{equation}
This value corresponds to inertial time $t_1=2\gamma^2_0v_0/g$, where
$\gamma_0$ is the Lorentz factor associated with $v_0$. At either
event
\begin{equation}\label{eq30}
\left.\frac{{\rm d}^2Z^2}{{\rm d}T^2}\right|_{Z=0}=-g(1-2v^2_0).
\end{equation}
On the other hand, $\dot{Z}(T=0)=v_0$ and $\dot{Z}(T=T_1)=-v_0$. That is, the free
particle ascends from $Z=0$ to $Z_0=(\gamma_0-1)/g$ at $T_0=T_1/2$ such that
$\tanh gT_0=v_0$, then descends back to $Z=0$ at $T_1$ and finally
approaches the horizon $Z=-1/g$ as $T\to \infty$. 
For $v_0=1/\sqrt{2}$, the particle has no acceleration at the
crossing events.

We note that for ultrarelativistic motion beyond the critical speed
$(v_0>1/\sqrt{2})$, the inertial acceleration of the free particle has
the opposite sign at crossing events compared to intuitive
expectations based on Newtonian mechanics; but, other aspects of
the motion are not affected. For instance, the critical speed does not
appear to play a role in the proper temporal intervals between the
crossing events. The relevant proper time of the Rindler observer
$T_1$ given by equation~\eqref{eq29} is  always less than the proper
time of the free particle $t_1/\gamma_0=2\gamma_0v_0$, since for
$0<v_0<1$,
\begin{equation}\label{eq31}
\ln \Big( \frac{1+v_0}{1-v_0}\Big) <\frac{2v_0}{\sqrt{1-v^2_0}}.
\end{equation}
That is, the length of the geodesic segment is maximum. 
This circumstance, usually called ``the twin paradox,''  is a
reflection of the absolute character of acceleration in the theory of
relativity.

For the limiting case of a null ray ($v_0=\pm 1$),
$P=1+gZ=\pm \dot{Z}=\exp (\pm gT)$ and
$\ddot{Z}=g\exp (\pm gT)$. The ray thus ascends monotonically toward
infinity or descends monotonically toward the horizon always with
positive acceleration. Moreover, the Fermi coordinate speed of the ray
can range from zero to infinity.

\section{Tidal effects}\label{s4}
We now turn our attention to motion in a gravitational field
\cite{1,2,3,4}. The spacetime metric in Fermi coordinates then takes the
form
\begin{eqnarray}
\label{eq32} g_{00}=-P^2+Q^2- \fr_{0i0j}X^iX^j+O(|{\bf
X}|^3),\\
\label{eq33} g_{0i} = Q_i -\frac{2}{3}\; 
\fr_{0jik}X^jX^k+O(|{\bf X}|^3),\\
\label{eq34} g_{ij}=\delta_{ij}-\frac{1}{3} \fr_{ikjl}X^kX^l +O
(|{\bf X}|^3),\end{eqnarray}
where
\begin{equation}\label{eq35}
\fr_{\alpha \beta \gamma \delta}(T)=R_{\mu\nu\rho \sigma}
\lambda^\mu_{\;\;(\alpha)} \lambda^\nu_{\;\; (\beta)}
\lambda^\rho_{\;\;(\gamma)} \lambda^\sigma _{\;\;(\delta)}
\end{equation}
is the projection of the Riemann tensor on the tetrad frame of the
observer.

To simplify matters, we concentrate on tidal effects only and assume
that the observer follows a geodesic along which a Fermi
coordinate system is constructed based on a parallel-propagated 
tetrad frame. Moreover, we neglect external forces on the
test particle. In this case, the equations of motion of the free
particle in the Fermi system, i.e.\ equations \eqref{eq7}--\eqref{eq10},
reduce to
\begin{eqnarray}\label{eq36}
\fl\frac{d^2X^i}{dT^2}+\fr_{0i0j}X^j+2 \fr_{ikj0}V^kX^j
\nonumber\\
&&\hspace{-1.8in}+\left(2\fr_{0kj0}V^iV^k+\frac{2}{3} \fr_{ikj\ell}
V^kV^\ell +\frac{2}{3} \fr_{0kj\ell} V^iV^kV^\ell \right)
X^j+O(|{\bf X}|^2)=0,
\end{eqnarray}
and
\begin{eqnarray}\label{eq37}
\nonumber \frac{1}{\Gamma^2}&=& 1-V^2+\fr_{0i0j}X^iX^j 
+ \frac{4}{3} \fr_{0jik}X^jV^i X^k\\
&&{}+ \frac{1}{3} \fr_{ikj\ell} V^iX^kV^jX^\ell +O(|{\bf X}|^3)\geq 0.
\end{eqnarray}
Equality holds for null rays $(\Gamma =\infty)$. In this case
the right-hand side of equation~\eqref{eq37} is a first integral
of the differential equation~\eqref{eq36}
and higher-order tidal terms cannot be neglected. The Fermi coordinate
system is admissible within a cylindrical region along the observer's
worldline with $|{\bf X}|<\mathcal{R}$, where $\mathcal{R}^{-2}(T)$ is
the supremum of $|\fr_{\alpha \beta \gamma \delta}(T)|$.  
Equation~\eqref{eq36} is the 
\emph{geodesic deviation equation} in Fermi coordinates, 
since it represents the relative motion of the free particle
 with respect to the fiducial observer that has been assumed here
to follow a geodesic.

To illustrate the tidal effects for test particles and null rays, we consider
the gravitational field of a Kerr black hole. The Kerr metric, 
in Boyer-Lindquist coordinates $(t,r,\vartheta ,\phi)$, is given by
\begin{eqnarray} \label{eq38}
-ds^2 &=-dt^2+\Sigma \big(
\frac{1}{\Delta}\,dr^2+d\vartheta ^2\big) + (r^2+a^2)\sin ^2\vartheta\,
d\phi^2\nonumber \\
&\quad + 2GM\frac{r}{\Sigma}(dt-a \sin^2 \vartheta\, d\phi)^2,
\end{eqnarray}
where $\Sigma =r^2+a^2\cos^2\vartheta$ and $\Delta =r^2-2GMr+a^2$. The
Kerr source has mass $M$ and angular momentum $J=Ma$. 
For a black hole, the specific angular momentum $a$
is such that $0\le a \le  GM$. The observer is
taken to be on an escape trajectory along the axis of rotation: It
starts at $\tau =0$ with $r_0>\sqrt{3}\, a$ on this axis and moves
radially outward reaching infinity with zero kinetic energy. Its
geodesic worldline is given by
\begin{equation}\label{eq39}
\frac{dt}{d\tau }=\frac{r^2+a^2}{r^2-2GMr+a^2},\quad \frac{dr}{d\tau
}=\sqrt{\frac{2GMr}{r^2+a^2}}.
\end{equation}

To determine Fermi coordinates, we   
choose a nonrotating orthonormal tetrad frame along the observer's worldline
such that $\lambda^\mu_{\;\;(3)}$ is parallel to the $z$-axis. 
In $(t,r,\vartheta ,\phi)$ coordinates, we have
\begin{equation}\label{eq40}
\lambda^\mu_{\;\;(0)} =(\dot{t}, \dot{r}, 0,0),\quad
\lambda^\mu_{\;\;(3)}=(\dot{t}\dot{r}, 1, 0, 0),
\end{equation}
where $\dot{t}=dt/d\tau$ and $\dot{r}=dr/d\tau $ are given in
display~\eqref{eq39}. Using the axial symmetry of the spacetime,
$\lambda^\mu_{\;\; (1)}$ and $\lambda^\mu_{\;\; (2)}$ can be chosen
uniquely up to a rotation about the $z$-axis. Once a pair is chosen at
$\tau =0$, then $\lambda^\mu_{\;\;(1)}$ and $\lambda^\mu_{\;\;(2)}$
are parallel propagated along the reference worldline. We do not
require the explicit transformation from Boyer-Lindquist to Fermi coordinates
to obtain the curvature components that appear in  
the approximate equations of motion~\eqref{eq36}--\eqref{eq37}.  
Indeed, the symmetries of the Riemann tensor make it
possible to represent these quantities as elements of a symmetric
$6\times 6$ matrix with indices that range over the set $\{ 01, 02,
03, 23,31,12\}$. The result is
\begin{equation}\label{eq41}
\left[\begin{array}{cc} E & B\\ B & -E\end{array}\right],
\end{equation}
where $E$ and $B$ are $3\times 3$ matrices that are symmetric and
traceless and correspond respectively to the electric and magnetic
parts of the Riemann tensor. Moreover, $E=\diag (-k/2,-k/2, k)$ and
$B=\diag (-q/2, -q/2, q)$, where
\begin{eqnarray}\label{eq42}
k=-2GM \frac{r(r^2-3a^2)}{(r^2+a^2)^3},\\
\label{eq43} q=2GMa \frac{3r^2-a^2}{(r^2+a^2)^3}.
\end{eqnarray}
The curvature components are computed along the worldline of the
observer, i.e.\ at $(T,{\bf 0})$ in Fermi coordinates. The electric and
magnetic curvatures in \eqref{eq42} and \eqref{eq43}, respectively,
only depend upon $r$. Therefore, it suffices to integrate the equation
for $\dot{r}$ in \eqref{eq39} to obtain $r(\tau )$, which when
substituted in \eqref{eq42} and \eqref{eq43} with $\tau \to T$ results
in $k(T)<0$ and $q(T)>0$, respectively. The equation of motion
\eqref{eq36} can thus be written as the system 
\begin{eqnarray}\label{eq44}
\ddot{X}&-\frac{1}{2} kX\left( 1-2
\dot{X}^2+\frac{4}{3}\dot{Y}^2-\frac{2}{3} \dot{Z} ^2\right)
+\frac{1}{3}k\dot{X} (5 Y\dot{Y} -7Z\dot{Z} )\nonumber\\
& +q[\dot{X} \dot{Y} \dot{Z} X-\dot{Z} Y (1+\dot{X}^2)-2\dot{Y}
Z]=0,\\
\label{eq45} \ddot{Y} &-\frac{1}{2}kY \left( 1-2\dot{Y}^2+\frac{4}{3}
\dot{X}^2- \frac{2}{3} \dot{Z}^2\right) +\frac{1}{3} k\dot{Y}
(5X\dot{X}-7Z\dot{Z})\nonumber \\
& -q [\dot{X} \dot{Y} \dot{Z} Y-\dot{Z} X (1+\dot{Y}^2) -2\dot{X}
Z]=0,\\
\label{eq46} \ddot{Z} &+kZ \left[ 1-2\dot{Z} ^2+\frac{1}{3} (\dot{X}^2
+\dot{Y}^2)\right] +\frac{2}{3} k\dot{Z} (X\dot{X} +Y\dot{Y}
)\nonumber\\
& -q (X\dot{Y} -\dot{X} Y)(1-\dot{Z}^2)=0,
\end{eqnarray}
where we have neglected higher-order tidal terms for the sake of
simplicity. These generalized Jacobi equations reduce to the standard
Jacobi equations when the relative motion is so slow that the velocity
terms can be neglected.
\begin{figure}
\vspace{1in}
\centerline{\epsfig{file=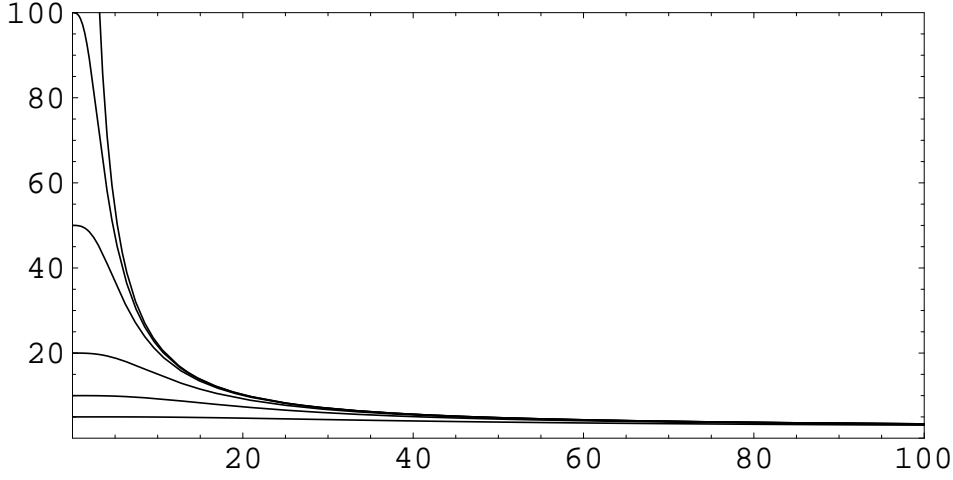,width=30pc}}
\caption{Graphs of $\Gamma_\parallel$ defined in~\eqref{eq48}
versus $T/(GM)$ 
approximated by
numerical integration of equation~\eqref{eq47} with initial $\Gamma_\parallel$
equal to 5, 10, 20, 50, 100 and 1000. We assume that
at $T=0$, $r_0=20 \, GM$ and $Z(0)=0$. In these graphs $a=GM$; however,
the corresponding graphs with $a=0$ are indistinguishable from those given
here.\label{fig:1}}
\end{figure}
\begin{figure}
\vspace{1in}
\centerline{\epsfig{file=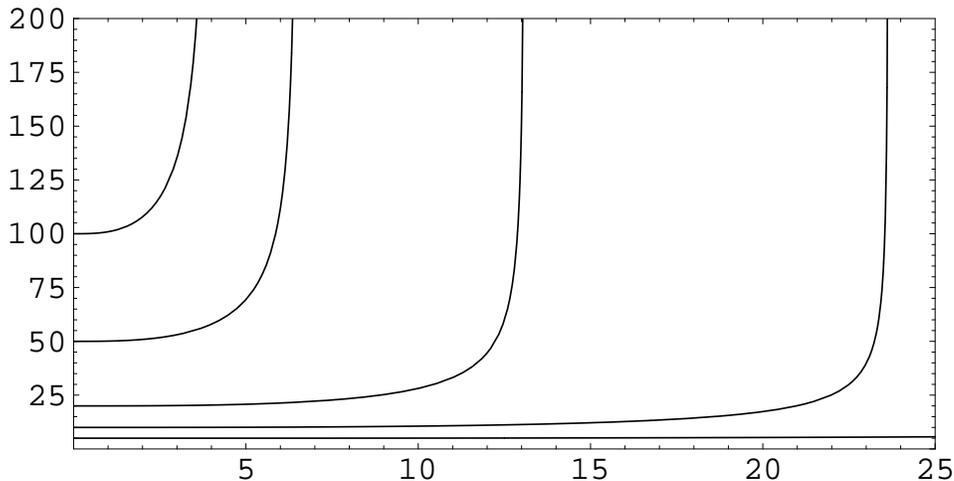,width=30pc}}
\caption{Graphs of $\Gamma_\perp$ defined in~\eqref{eq50} versus
$T/(GM)$
approximated by
numerical integration of equation~\eqref{eq49} with $a=GM$ and
initial $\Gamma_\perp$
equal to 5, 10, 20, 50 and 100.   We assume that
at $T=0$, $r_0=20 \,GM$ and $X(0)=0$. Note that $\Gamma_\perp$
diverges at $T/(GM)\approx$ 23.7, 13.1, 6.5 and 4.0 for the initial
Lorentz factors $\Gamma_\perp(0)=$ 10, 20, 50 and 100,
respectively \label{fig:2}}
\end{figure}

Consider, for instance, the one-dimensional motion of the test
particle along the $z$-direction. Equations \eqref{eq44}-\eqref{eq46}
with $X(T)=Y(T)=0$ for $T\geq 0$ reduce to
\begin{equation}\label{eq47}
\frac{d^2Z}{dT^2}+k(1-2\dot{Z}^2)Z=0.
\end{equation}
The behavior of the solutions of this equation, which contains the
critical speed $V_c=1/\sqrt{2}$, has been studied in detail in our
previous work \cite{5}.
\begin{figure}
\vspace{1in}
\centerline{\epsfig{file=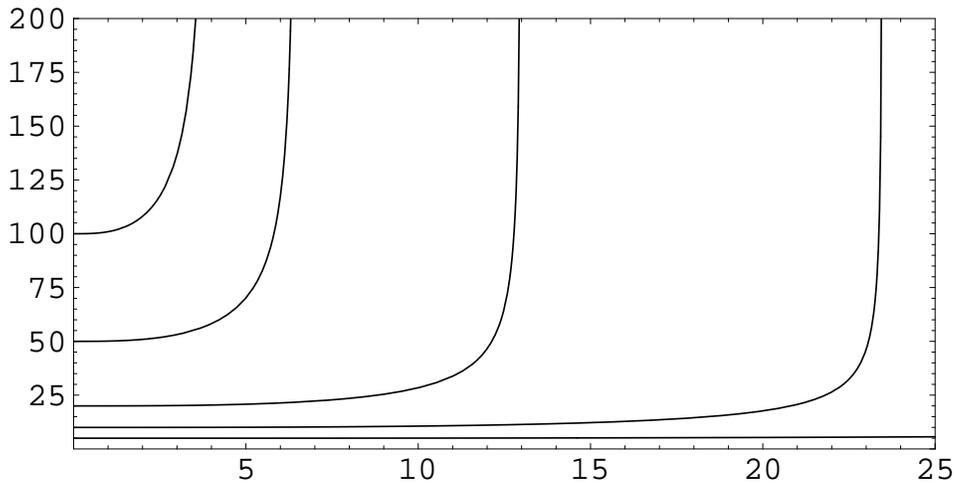,width=30pc}}
\caption{Graphs of $\Gamma_\perp$ defined in~\eqref{eq50}
versus $T/(GM)$
approximated by
numerical integration of equation~\eqref{eq49} with $a=0$ and
 initial $\Gamma_\perp$
equal to 5, 10, 20, 50 and 100. We assume that
at $T=0$, $r_0=20\,GM$ and $X(0)=0$. Note that $\Gamma_\perp$
diverges at $T/(GM)\approx$ 23.5, 13.0, 6.4 and 3.9 for the
initial Lorentz factors $\Gamma_\perp(0)=$ 10, 20, 50 and 100,
respectively.  \label{fig:3}}
\end{figure}
Suppose that $Z(T=0)=0$ and $\dot{Z}(T=0)>1/\sqrt{2}$; then, the
particle starts from the observer's location and decelerates along the
$z$-axis toward the critical speed $1/\sqrt{2}$. In this case $\Gamma$
given by equation \eqref{eq37} reduces to
\begin{equation}\label{eq48}
\Gamma_\parallel =\frac{1}{\sqrt{1-\dot{Z}^2+kZ^2}},
\end{equation}
whose behavior is depicted in figure~\ref{fig:1}.
This extends our previous results~\cite{5} and is consistent
with the rapid decrease of curvature $k$ away from the black hole.

A more interesting situation arises for motion normal to the direction
of rotation of the black hole. Equations \eqref{eq44}--\eqref{eq46}
with $Y(T)=Z(T)=0$ for $T\geq 0$ reduce to 
\begin{equation}\label{eq49}
\frac{d^2X}{dT^2} -\frac{1}{2} k(1-2\dot{X}^2) X=0,
\end{equation}
which turns out to be the general equation for one-dimensional radial
motion perpendicular to the rotation axis of the black hole as a
consequence of the axial symmetry of the configuration under
consideration here. The corresponding Lorentz factor is given by
\eqref{eq37} as
\begin{equation}\label{eq50}
\Gamma_\bot =\frac{1}{\sqrt{1-\dot{X}^2-\frac{1}{2}kX^2}} .
\end{equation}
We note that for ultrarelativistic motion, the particles gain tidal
energy and reach the speed of light as in figure~\ref{fig:2} for
an extreme Kerr black hole $(a=GM)$ and in figure~\ref{fig:3} for
a Schwarzschild black hole $(a=0)$. It follows from a comparison of 
figures~\ref{fig:2} and~\ref{fig:3} that the tidal acceleration 
mechanism does not depend sensitively 
upon the specific angular momentum of the black hole;
in fact, it appears that the time it takes for $\Gamma_\perp$
to diverge is in general slightly longer for $a=GM$ as compared
to the $a=0$ case. 
Moreover, the divergence of $\Gamma_\perp$ 
indicates that our test-particle approach breaks down.

In two recent papers~\cite{new6}, we approximated $\Gamma_\perp$ by 
$( 1 - \dot X^2 ) ^ {-1/2}$
and plotted its approach
toward infinity as a consequence of the ultrarelativistic tidal 
acceleration mechanism. The present calculation of $\Gamma_\perp$ 
is based on an improved approximation scheme 
and goes beyond our previous work. 
To reach the speed of light --- or, equivalently, 
for the timelike particle worldline to become null --- the Fermi 
speed $\dot X$ should reach a value that is greater than unity as 
$k$ in equation~\eqref{eq50} is negative. 
We expect that this value of $\dot X$ is reached at a later time $T$
as demonstrated by a comparison of figures~\ref{fig:2} and~\ref{fig:3}  
with our previous results~\cite{new6}. Even the present calculations 
are approximate as they are based on dropping the higher-order tidal terms. 
It is clear that at a sufficiently long time $T \sim \mathcal{R}$, 
the Fermi coordinate system breaks down. 
The kinematic breakdown of Fermi coordinates, i.e.\ their inadmissibility, 
is logically independent of the divergence of $\Gamma_\perp$, 
which indicates the failure of the dynamical theory presented here. 
Our test-particle approximation ignores any back reaction; 
furthermore, for charged particles the electromagnetic field 
configuration near the black hole could enhance or hinder the 
tidal acceleration mechanism. In any case, this breakdown of
our dynamics nevertheless 
implies that sufficiently energetic particles emerge from the vicinity 
of the black hole after having experienced tidal acceleration 
by the gravitational field of the collapsed source. 
In this way, highly energetic particles may be created by 
microquasars in our galaxy. It is important to emphasize 
that the tidal acceleration mechanism is independent of the 
horizon structure of the black hole. 
These results are interesting in connection
with the origin of the highest energy cosmic rays since cosmic rays 
with energies above the Greisen-Zatsepin-Kuzmin limit 
$( \sim 10 ^{20}\,{\rm eV} )$ 
are not expected to reach the Earth from distant galaxies \cite{15}.

\section{Discussion}\label{s5}
Fermi coordinates constitute a geodesic coordinate system that is a
natural extension of the standard Cartesian system and is
indispensable for the theory of measurement in relativistic
physics. We have discussed the general equation of motion of a
pointlike test particle, as well as the limiting case of a null ray,
in Fermi coordinates. Inertial and tidal effects of ultrarelativistic
particles with speeds above the critical speed $c/\sqrt{2} \simeq
0.7c$ have been emphasized. This work goes beyond our previous work in
this direction and strengthens the basis for our results and
conclusions \cite{5,new6,6}.

\appendix
\section{Null rays}\label{sA}

Let us compute the effective external force per unit mass given by
equation \eqref{eq21} for the electromagnetic case, namely,
\begin{equation}\label{eqA1}
A^\mu =\frac{\hat q}{m}F^\mu_{\;\;\nu} \frac{dx^\nu}{ds}
\end{equation}
in accordance with the Lorentz force law for the motion of a particle
of charge $\hat q$ in an external field $F_{\mu\nu}$. Neglecting
radiation reaction, we find that
\begin{equation}\label{eqA2}
\mathcal{F}^i=\hat q\left( \frac{d\lambda}{dT}\right) (F^i_{\;\;
0}+F^i_{\;\; j} V^j-F^0_{\;\; j} V^iV^j),
\end{equation}
where $\lambda$ is the affine parameter defined by $ds= m\,d\lambda$.

A massless charged particle does not exist; therefore, we assume that
$\hat q\to 0$ as $m\to 0$. With this assumption, it turns out that the
massless limit of the trajectory of a charged particle is a null
geodesic.

There exist nongeodesic null curves in Minkowski spacetime. These may
be physically interpreted as follows: Imagine the path of a null
electromagnetic ray that is reflected from a collection of mirrors
fixed in different places in space. The path consists of 
straight (i.e.~geodesic) null segments
separated by mirrors. Next imagine a limiting situation involving an
infinite number of such idealized pointlike mirrors. The resulting smooth curve
is a nongeodesic null curve. An example of such a curve is
\begin{equation}\label{eqA3}
x^\mu (\lambda)=(\alpha \lambda ,\alpha \sin \theta \sin \lambda
,\alpha \sin \theta \cos \lambda , \alpha \lambda \cos \theta ),
\end{equation}
where $\alpha$ and $\theta$ are constants. The generalization to
curved spacetime is straightforward.

\section*{Acknowledgments}

BM is grateful to Donato Bini and Robert Jantzen for helpful correspondence.


\section*{References} 
\begin{thebibliography}{xxxxx}
\bibitem{1} Synge J L 1960 
\emph{Relativity: The General Theory}, North-Holland, Amsterdam
\bibitem{2} Marzlin K-P 1994 {\it Gen. Rel. Grav.} {\bf 26} 619\\
Marzlin K-P 1994 {\it Phys. Rev. D} {\bf 50} 888\\
Marzlin K-P 1996 {\it Phys. Lett A} {\bf 215} 1
\bibitem{3} Mashhoon B 1975 {\it Astrophys. J.} {\bf 197} 705\\
Mashhoon B 1977
{\it Astrophys. J.} {\bf 216} 591
\bibitem{4} Ni W-T and Zimmermann M 1978
{\it Phys. Rev. D} {\bf 17} 1473\\
 Li W-Q and  Ni W-T 1979
{\it J. Math. Phys.} {\bf 20} 1473\\
Li W-Q and Ni W-T 1979 {\it J. Math. Phys.} {\bf 20} 1925 
\bibitem{5} Chicone C and  Mashhoon B 2002
{\it Class. Quantum Grav.} {\bf 19} 4231\\
Chicone C and  Mashhoon B 2004
{\it Int. J. Mod. Phys. D} {\bf 13} 945
\bibitem{new6} 
 Chicone C  and  Mashhoon B 2004
{\it Preprint} astro-ph/0404170\\
Chicone C  and  Mashhoon B 2004
{\it Preprint} astro-ph/0406005
\bibitem{6} Chicone C and Mashhoon B 2004 \emph{Class. Quantum Grav.}, in press
\bibitem{7}  Mashhoon B 2003 
in: {\it Advances in General Relativity and Cosmology}, edited by G.
Ferrarese, Pitagora, Bologna, pp. 323--334
\bibitem{new} de Felice F 1995 \emph{Phys. Rev. A} {\bf 52} 3452
\bibitem{8} Estabrook F B and Wahlquist H D 1964 
{\it J. Math. Phys.} {\bf 5} 1629
\bibitem{9} DeFacio B, Dennis P W and Retzloff D G 1978
{\it Phys. Rev. D} {\bf 18} 2813\\
 DeFacio B, Dennis P W and Retzloff D G 1979 {\it Phys. Rev. D} {\bf 20} 570
\bibitem{10} Jaffe J and Shapiro I I 1972 {\it Phys. Rev. D} {\bf 6} 405 
\bibitem{11} Jantzen R T, Carini P and Bini D 1992 
{\it Ann. Phys.} ({\it  N Y} ) {\bf 215} 1\\
Bini D, Carini P and Jantzen R T 1995 {\it Class. Quantum Grav.} {\bf
12} 2549
\bibitem{12} Moliner I, Portilla M and Vives O 1995 {\it Phys. Rev. D}
{\bf 52} 1302
\bibitem{13} Bunchaft F and Carneiro S 1998 {\it Class. Quantum Grav.}
{\bf 15} 1557
\bibitem{14} Rindler W 1969 {\it Am. J. Phys.} {\bf 34} 1174\\
Rindler W 2001 {\it Relativity: Special, General and Cosmological},
Oxford University Press, Oxford
\bibitem{15} Anchordoqui L A, Dermer C D and Ringwald A 2004 {\it
Preprint} hep-ph/0403001

\end{thebibliography}
\end{document}